\documentclass[%
 reprint,
superscriptaddress,
 amsmath,amssymb,
 aps,
 prl,
]{revtex4-2}
\usepackage[T1]{fontenc}
\usepackage{lmodern}
\usepackage[utf8]{inputenc}
\usepackage[english]{babel}
\usepackage{graphicx}
\usepackage{dcolumn}
\usepackage{bm}     
\usepackage{hyperref}
\usepackage{xcolor}
\usepackage{soul}
\hypersetup{colorlinks=true,citecolor={blue},linkcolor={blue},urlcolor={blue}}

\begin{document}

\title{Emission enhanced exciton-polariton condensates with optical feedback}
\author{R.~Mirek}
\author{M.~Furman}
\affiliation{Institute of Experimental Physics, Faculty of Physics, University of Warsaw, ul. Pasteura 5, PL-02-093 Warsaw, Poland}
\author{A.~Opala}
\affiliation{Institute of Experimental Physics, Faculty of Physics, University of Warsaw, ul. Pasteura 5, PL-02-093 Warsaw, Poland}
\affiliation{Institute of Physics, Polish Academy of Sciences,\\ Aleja Lotnik\'ow 32/46, PL-02-668 Warsaw, Poland}
\author{M.~Kr\'ol}
\author{W.~Pacuski}
\affiliation{Institute of Experimental Physics, Faculty of Physics, University of Warsaw, ul. Pasteura 5, PL-02-093 Warsaw, Poland}
\author{J.~Szczytko}
\affiliation{Institute of Experimental Physics, Faculty of Physics, University of Warsaw, ul. Pasteura 5, PL-02-093 Warsaw, Poland}
\author{H.~Sigurðsson}
\affiliation{Institute of Experimental Physics, Faculty of Physics, University of Warsaw, ul. Pasteura 5, PL-02-093 Warsaw, Poland}
\affiliation{Science Institute, University of Iceland, Dunhagi 3, IS-107, Reykjavik, Iceland}
\author{B.~Pi\k{e}tka}
\email{barbara.pietka@fuw.edu.pl}
\affiliation{Institute of Experimental Physics, Faculty of Physics, University of Warsaw, ul. Pasteura 5, PL-02-093 Warsaw, Poland}

\begin{abstract}
Optical feedback is a well-known method of controlling laser dynamics, which has been widely studied in photonic systems to induce complex behaviors such as chaos or enhanced coherence. However, its application to systems in the strong light-matter coupling regime remains unexplored. 
In this work, we introduce a delayed optical feedback loop into a nonresonantly pumped polariton condensate. By feeding part of the emission
back into the cavity to seed the next condensate, we observe a strong increase in the output intensity, up to 110\%. We explain this effect using a classical rate equation model for the condensate coupled to excitonic reservoirs. Our results evidence that polariton condensates can respond strongly to optical feedback, congruent with well known polariton amplification techniques using resonant pump-probe setups. Our method opens new possibilities for using polariton feedback to connect multiple condensates and can be an essential step toward neuromorphic computing based on recurrent signaling in photonic systems.
\end{abstract}
\maketitle

\section{Introduction}

Delayed optical feedback in semiconductor lasers has been studied for a long time~\cite{Lang_IEEE1980, Soriano_RevModPhys2013} giving rise to chaotic dynamics with practical advantages in optical-based communications and random bit generation~\cite{Sciamanna_NatPhot2015}. Within the field of quantum optics, it has been shown that feedback can dramatically affect the photon statistics~\cite{Albert_NatComm2011} and mode-switching dynamics~\cite{Holzinger_SciRep2019}.
However, semiconductor microcavities operating in the strong light-matter coupling regime have been scarcely explored in the context of optical feedback. This is particularly surprising since the associated light-matter quasiparticles, exciton-polaritons, are known for large optical nonlinearities, ultrafast (picosecond) responses towards external fields, and strong parametric amplification~\cite{Ciuti_SemSciTech2003, Baumberg_PSSb2005}. These polaritonic qualities have sparked intense research efforts towards implementing microcavity systems as elements in compact optical-based information processing technologies~\cite{Kavokin_NatRevPhys2022}. Here, we demonstrate that optical feedback in the strong coupling regime can significantly affect the coherent emission intensity coming from a condensate of polaritons.

Exciton-polaritons are composite bosons that form under strong light-matter interaction between quantum well excitons and confined cavity photons (from here on just \textit{polaritons}). The photonic component imbues polaritons with small effective mass and easy optical detection, whereas their exciton component allows them to strongly interact with each other due to repulsive Coulomb interactions. Because of the hybrid light-matter composition of polaritons, they possess exciting ingredients relevant for on-chip optical technologies such as angle-resonant parametric amplification~\cite{Savvidis_PRL2000, Huang_PRB2000, Houdre_PRL2000, Ciuti_SemSciTech2003, Baumberg_PSSb2005, Zhao_NatNanoTech2022}, optical switching~\cite{Amo_NatPhot2010, dreismann2016sub, Feng_SciAdv2021, zasedatelev2021single, suarez2021ultrafast, chen2022optically, Jianbo_ACS2023}, and transistor operation~\cite{Gao_PRB2012, anton2012dynamics, Ballarini_NatComm2013, Zasedatelev_NatPhot2019}, all the way up to room temperature in high exciton binding-energy materials~\cite{Zhao_NatNanoTech2022, zasedatelev2021single, Zasedatelev_NatPhot2019, chen2022optically}.  

Before nonequilibrium condensation of polaritons became well established~\cite{Byrnes_NatPhys2014}, the most prominent experiments generating coherent polariton emission were angle-resolved pump-probe experiments based on parametric amplification~\cite{Savvidis_PRL2000}. Reviews on the subject can be found in Refs.~\cite{Ciuti_SemSciTech2003, Baumberg_PSSb2005}. In these experiments, the excitation is a resonant and narrow linewidth laser which offered greater control than nonresonant excitation schemes~\cite{Dang_PRL1998, Senellart_PRB2000}. Polaritons could be pumped at certain energies and wavevectors~\cite{Savvidis_PRL2000} to then undergo triggered stimulated scattering into a macroscopically occupied state when a probe beam was added. More recently, parametric amplification of polaritons was demonstrated all the way up to room temperature by using more indirect nonresonant excitation schemes. There, an incoherent laser photoexcites a reservoir of ``hot'' excitons which can be triggered into a final state through e.g. vibronic relaxation channels in cavities with organic compounds~\cite{zasedatelev2021single}, or by intersubband parametric scattering in thick perovskites cavities~\cite{Wu_NanoLett2021} and one-dimensional ZnO whispering gallery cavities ~\cite{Xie_PRL2012, chen2022optically}.

The abovementioned strategies rely on using more than one laser to trigger strong optical response from the system. However, vivid enhancement of the cavity emission from a polariton condensate is also possible by using only a single nonresonant laser (i.e., pump), which creates a reservoir of uncondensed particles, and then feeding back into the cavity the emission from previous pump pulse to facilitate condensation. In this study we implement such a system by connecting the polariton cavity emission to a feedback loop which quasiresonantly seeds the polariton state for subsequent nonresonant excitaiton pulses. The results show a dramatic enhancement in the emission intensity of the cavity by as much as $~110\%$ attached to a sharp, several picosecond, resonant amplification peak. The experimental results are in good agreement with a classical rate-equation model describing a polariton condensate population responsible for the coherent light emission, coupled to populations of excitonic reservoirs.

\section{Results}
\subsection{Experimental setup}

The investigated system is a planar optical microcavity consisting of two distributed Bragg reflectors with 16 and 19 alternating (Cd,Zn,Mg)Te/(Cd,Mg)Te layers, respectively, separated by an approximately 600\,nm thick (Cd,Zn,Mg)Te layer, forming a microcavity of quality factor $Q = 300$. The microcavity contains
three pairs of quantum wells of 20\,nm with a small concentration (about 0.5\%) of manganese ions each. The structure was grown on a nontransparent (100)-oriented GaAs substrate by molecular beam epitaxy. 
 
\begin{figure}[t]
\centering
\includegraphics[width=\linewidth]{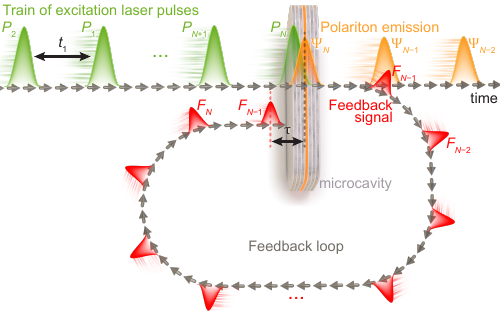}
\caption{Scheme of the experiment. Pulses from nonresonant pump laser (green) excite the microcavity leading to polariton emission (orange). Part of the emission (red) is redirected into the feedback loop and used as a resonant feedback for the subsequent laser pulse.}
\label{setup}
\end{figure}

Figure~\ref{setup} illustrates the concept of our experiment. The sample was excited nonresonantly with a pulsed laser 4\,ps of full-width-half-maximum. The interval between consecutive pulses was approximately $t_1 = 13$\,ns (76\,MHz repetition rate). The pump is represented by the green pulse signal in Figure~\ref{setup}. The laser was linearly polarized and its energy tuned to the first Bragg minimum on high energy side of the microcavity stop-band (nonresonant excitation). Every single laser pulse triggers polariton photoluminescence~(PL) from the microcavity (shown as the orange signal in Figure~\ref{setup}).  A fraction of this emission was then directed into a time-delayed feedback loop and subsequently used as a resonant seed for the emission from the next pulse (depicted as a small red pulse in Figure~\ref{setup}). The delay time $\tau$ of the emission pulse is defined as the difference between the propagation time in the feedback arm against the pump pulse repetition period $\tau = t_\text{FB} - t_1$. A negative delay $\tau<$ corresponds to a feedback seed arriving before the next pump pulse, and conversely $\tau>0$ means the signal arrives afterward. The zero delay $\tau=0$ was determined from interference measurements using laser light reflected from the sample. Further details on the experimental setup and determining the zero delay are presented in the Appendix. 

\begin{figure}
\centering
\includegraphics[width=\linewidth]{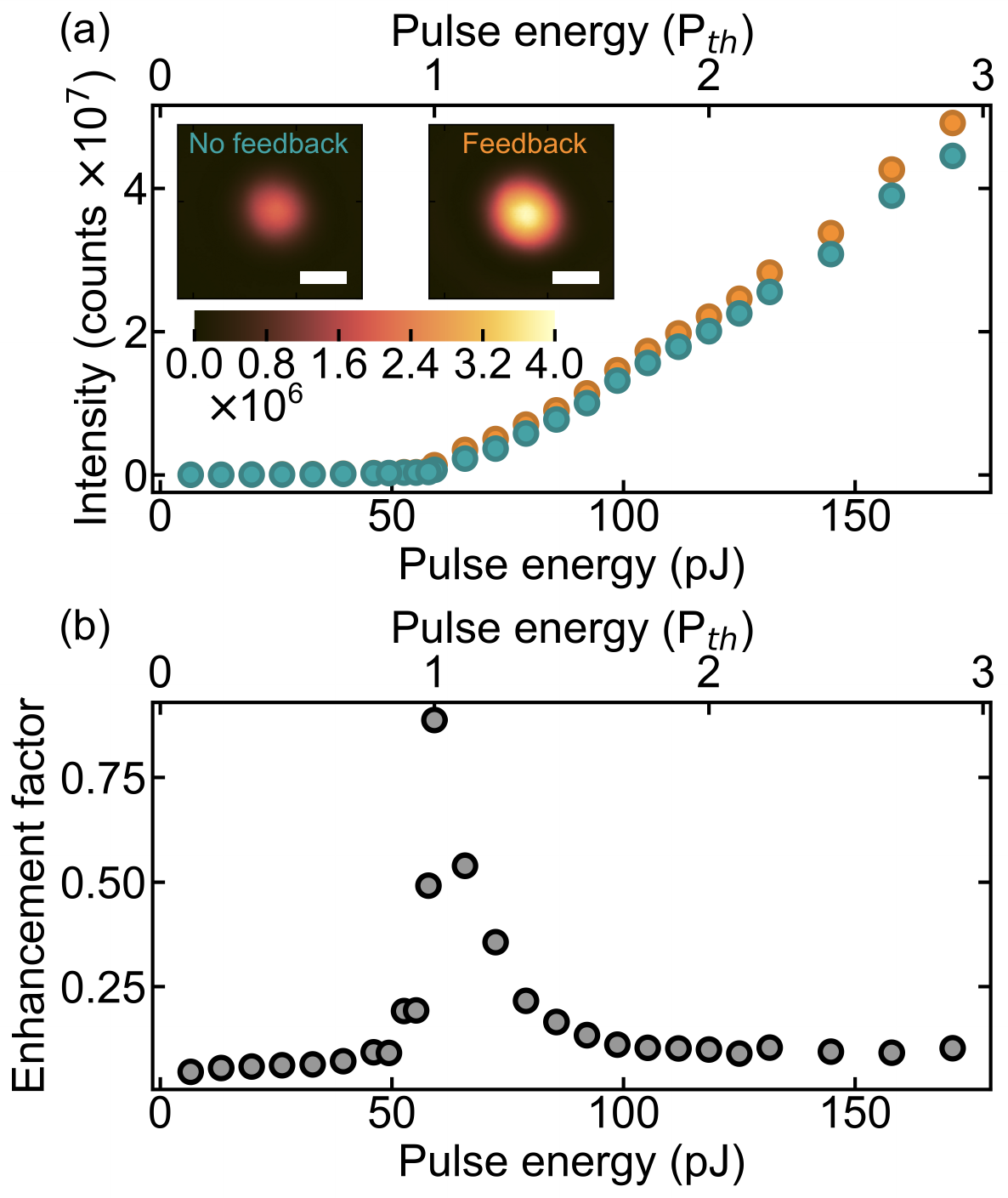}
\caption{Feedback enhancement of the cavity emission. (a) Microcavity input-output relationship with (orange dots) and without (cyan dots) feedback with delay time of $\tau = -4$\,ps. Insets show real space images of the cavity PL emission at pump pulse fluence of 59\,pJ without (left) and with (right) feedback. The white scale bars indicate 2\,$\mu$m. (b) Enhancement factor $\eta$ of the condensate as a function of the pump power (pulse energy) and $\tau = -4$\,ps time delay.}

\label{nonlinearities}
\end{figure}

\begin{figure*}
\centering
\includegraphics[width=\linewidth]{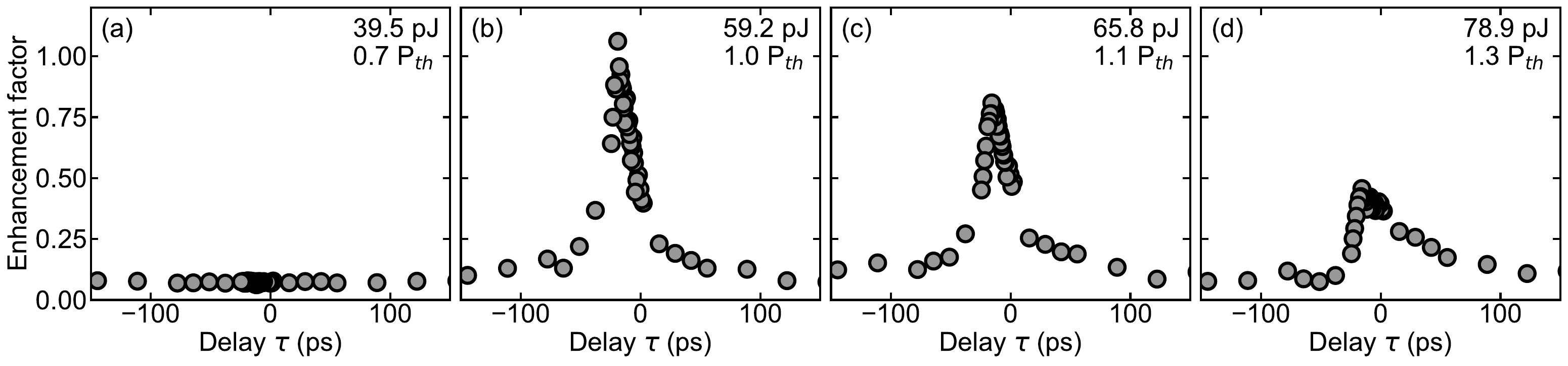}
\caption{Enhancement of the polariton PL as a function of feedback delay and increasing pumping power going from left to right.}
\label{feedback}
\end{figure*}

\subsection{Condensate feedback}

The integrated cavity PL as a function of pump power is shown in Fig.~\ref{nonlinearities}(a) with (orange dots) and without (cyan dots) feedback. In both cases, a sudden increase in the emission intensity occurs around $P_\text{th} \approx 55$\,pJ defining the polariton condensation threshold power (fluence). Here, the delay between the cavity emission re-entering the cavity and subsequent pump pulse was set to $\tau = -4$\,ps. Both curves show qualitatively the same behavior, but with higher cavity emission above the condensation threshold when feedback is present. The differences are the most prominent in the vicinity of the condensation threshold, what can be described as a lowering of the condensation threshold with the addition of the seed from the feedback arm.

To quantify the emission intensity difference with and without the additional resonant seed we define an enhancement factor (i.e. feedback loop gain or differential signal): 
\begin{equation}
    \label{eq:FEEDFeedback}
    \eta \equiv \frac{\langle I_{F}\rangle - \langle I\rangle}{\langle I\rangle},
\end{equation}
where $\langle I_{F}\rangle$ and $\langle I\rangle$ denote the integrated cavity PL with and without feedback, respectively. The enhancement is shown in Fig.~\ref{nonlinearities}(b) showing a clear resonant feature in the enhancement of the condensate emission as a function of excitation pump power. Maximum enhancement of $\approx 90\%$ appears around the threshold pump power $\approx 60$\,pJ (i.e., the threshold power without feedback). For pump energies below 50\,pJ and above 100\,pJ, we observed only a modest increase of the total intensity, with enhancement below 5$\%$. The insets in Fig.~\ref{nonlinearities}(b) show the spatially resolved cavity emission at nonresonant excitation energy of 59\,pJ without (left) and with (right) the feedback, respectively. The results evidence that light coming from the feedback line facilitates bosonic stimulation of the condensate by resonantly seeding the polariton state before the arrival of the nonresonant pump pulse. The pump pulse photoexcites charge-carriers which relax down to the exciton level to form a reservoir of incoherent excitons which in-turn undergo stimulated relaxation to form the polariton condensate~\cite{Byrnes_NatPhys2014}.

Figure~\ref{feedback} shows results on the condensate PL emission enhancement for varying time delay $\tau$ between the feedback signal and the pump pulse. Above the condensation threshold, a clear resonant feature can be observed indicating the coincidence of the signal and the pump pulse, seeding the polariton state and facilitating subsequent stimulation of the condensate. Figure~\ref{feedback}(a) shows the case of far below threshold with no change in enhancement. When pumped at threshold [Figure~\ref{feedback}(b)], an enhancement peak appears showing almost $\approx 110\%$ enhancement of the PL with a maximum located at $ \tau = - 18.6$\,ps. At such powers and short delays, the light passing through the feedback line is sufficient to trigger stimulated bosonic scattering from the high energy-momentum polaritons and excitons into a coherent condensate signal reminiscent of the well known polariton parametric amplifier~\cite{Savvidis_PRL2000, Stevenson_PRL2000}. The reason the enhancement is maximized at negative delays, implying a shorter feedback arm path length, is because of the finite build-up-time of the condensate when the last pump pulse hit the cavity. Indeed, it is safe to assume that maximum enhancement occurs when the feedback signal coincides with the pump pulse meaning that the time it took for the condensate to form and emit light was $t_\text{c} = 18.6$\,ps. This is a reasonable timescale for the ultrafast charge-carrier relaxation timescales into excitons and polaritons in semiconductor cavities.

For higher pump powers shown in Figs.~\ref{feedback}(c) and~\ref{feedback}(d) the peak enhancement drops. This is an expected result since the nonresonant pump can now by itself induce condensation and does not rely as much on the feedback to trigger stimulated scattering. Interestingly, we observe for increasing pump powers that the enhancement factor becomes more asymmetric about its peak with a steep rise followed by a slower fall-off towards positive delays. 

\begin{figure}
\centering
\includegraphics[width=\linewidth]{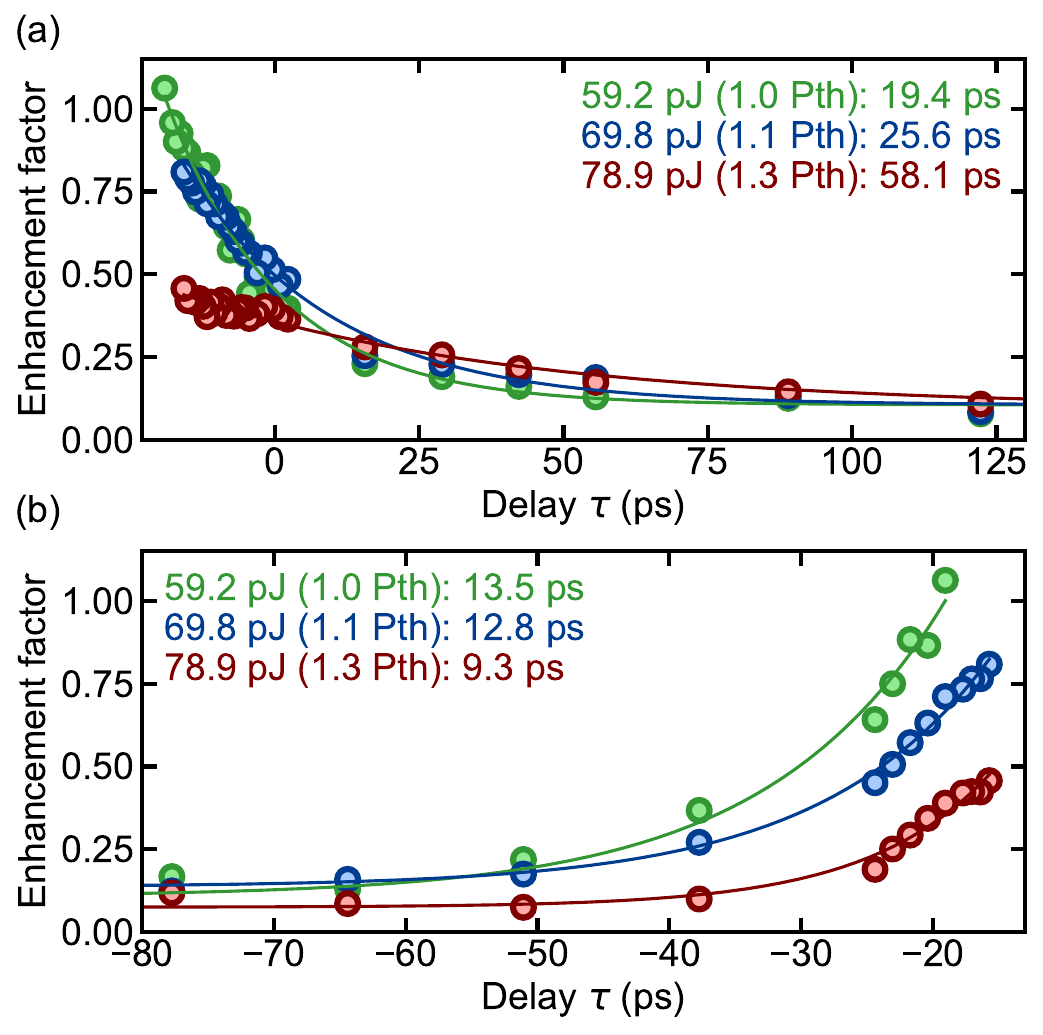}
\caption{(a) Decay and (b) build-up of the enhancement factor for different pump powers. The solid curves represent the exponential fit to the data, from which the decay/build-up times were obtained shown in the legend of each panel.}
\label{decay-build-up}
\end{figure}

The asymmetric \textit{decay} and \textit{build-up} of the enhancement factor as a function of time delay are shown in more detail in Fig.~\ref{decay-build-up}. We fitted the experimental data with exponential functions such to obtain their characteristic timescales depicted in the legends of each panel. Notably, the build-up time of the enhancement factor [Fig.~\ref{decay-build-up}(b)] shortened with increasing excitation power whereas its decay (fall-off) lengthened [Fig.~\ref{decay-build-up}(a)]. Moreover, the decay times of the enhancement are noticeably longer than the corresponding build-up times. We propose that this behavior is a manifestation of the longer exciton reservoir lifetimes compared to the shorter polariton condensate lifetimes (see Theory section). In other words, at positive delays, the exciton reservoir lives long enough to sense the feedback signal, whereas at negative delays the early arrival of the feedback signal triggers stimulation and depletes the reservoir fast.

\section{Theory}
We theoretically describe the enhancement of the polariton signal in the presence of feedback using a classical rate-equation model for the coherent condensate polaritons $\psi(t) = \langle \hat{\psi}(t) \rangle$ under scalar (no polarization) and single-mode ($k\approx 0$) approximations coupled to excitonic reservoirs. Our coupled rate equations account for the incoherent reservoir populations of excitons in two parts. First, an inactive reservoir $n_I$ composed of high-momentum excitons that do not directly scatter into the condensate, and an active reservoir $n_R$ describing excitons around the lower polariton bottleneck that can scatter into the condensate: 
\begin{align}
\label{eq:1}
i\frac{d \psi}{d t}  = & \ \left[\frac{i}{2}(Rn_R-\gamma_C)+U(t)\right]\psi\\ \notag
&+\alpha \psi(t-t_\text{FB})+\eta(t), \\
\label{eq:2}
\frac{d n_R}{d t} = & \ \kappa n_I^2-\gamma_Rn_R-Rn_R|\psi|^2, \\
\label{eq:3}
\frac{d n_I}{d t} = & \ \sum_{n=1}^N P_0 e^{-(t - n \cdot t_1)^2/2\sigma_t^2} -\kappa n_I^2-\gamma_In_I.
\end{align}
The feedback time was defined earlier as $t_\text{FB} = t_1 + \tau$ where $t_1$ is the repetition period of the pump pulses and $\tau$ the delay parameter. The rate of stimulated scattering from the bottleneck reservoir into the condensate is given by $R = 0.005$\,ps$^{-1}$, the feedback strength is denoted by $\alpha=0.05$, and $\eta(t)$ is a Langevin white noise operator $\langle \eta(t) \eta(t') \rangle = \bar{\eta} P_0 \delta(t-t')$ with amplitude $\bar{\eta}P_0$ describing spontaneous scattering from the reservoir~\cite{Ciuti_SemSciTech2003}. The potential
\begin{equation}
U(t)=g_C|\psi|^2+2g_R(n_R+ n_I),
\end{equation}
describes an energy shift coming from repulsive interactions between polaritons in the condensates $g_C = g_0 |X|^4 / (2 N_\text{QW})$ and from polaritons interacting with reservoir excitons $g_R = g_0 |X|^2 / (2 N_\text{QW})$. Here, $\hbar g_0 = 5$ $\mu$eV $\mu$m$^{2}$ is the exciton-exciton interaction constant, $|X|^2 = 0.5$ is the exciton Hopfield coefficient of the polariton at zero exciton-photon detuning, and $N_\text{QW}=6$ is the number of quantum wells in the cavity~\cite{Mirek_PhysRevB}. The factor $1/2$ appears because the triplet (same exciton spin) interaction term is dominant and our excitation pump, being linearly polarized, creates equal amounts of spin-up and spin-down excitons. Parameters $\gamma_C=1$\,ps$^{-1}$, $\gamma_R = \gamma_C/5$ and $\gamma_I = \gamma_C/100$ are respectively the decay rates of polaritons and active and inactive reservoir excitons~\cite{time-delayed-effects}. The conversion rate of inactive excitons into active excitons is denoted by $\kappa = 0.05$\,ps$^{-1}$, and nonresonant pulsed excitation is denoted by the first term in~\eqref{eq:3} of amplitude $P_0$ and width $\sigma_t$. 

We solve the system of stochastic differential equations numerically using the Milstein method over multiple random realizations of the white noise term to obtain average behavior of the condensate. For practical purposes, we do not simulate the full $t_1 = 12$\,ns gap between pump pulses given by the sum in~\eqref{eq:3}. Instead, we only simulate one pulse over a sufficiently long period where the reservoirs and condensate are populated (e.g. $<100$ ps interval). We then repeat the simulation for the next pulse but now with $\alpha \psi(t - t_\text{FB})$ replaced by $\alpha \tilde{\psi}(t - \tau)$ where $\tilde{\psi}$ represents the condensate from the previous simulation entering with delay $\tau$. This cuts down redundant computation time. After about $N > 10$ iterations (pulses) the system has converged. 

\begin{figure}
    \centering
    \includegraphics[width=\linewidth]{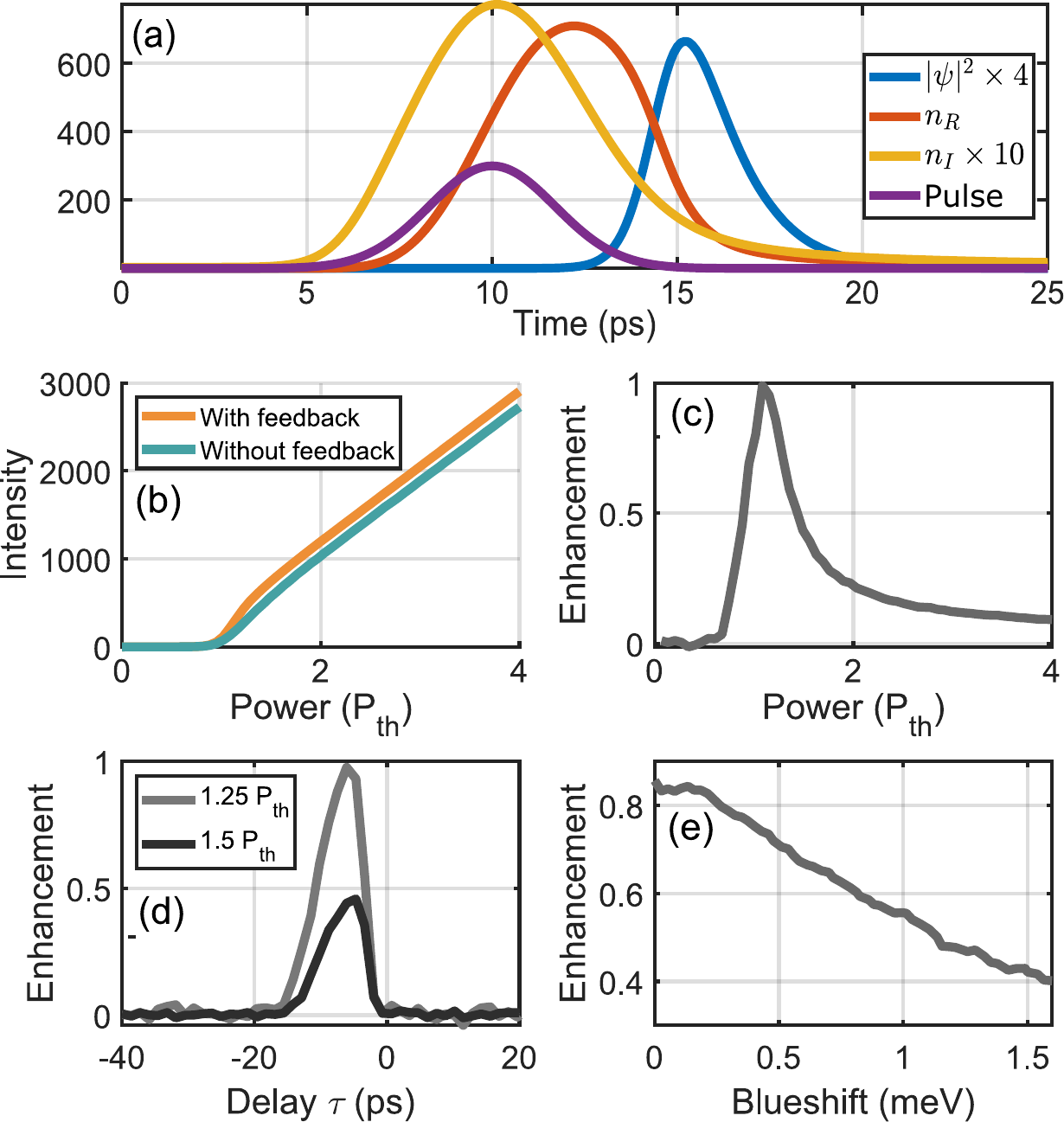}
    \caption{Simulations of the experiment. (a) An example single pulse behavior of the system. (b) Integrated intensity and (c) enhancement $\eta$ of the condensate $|\psi|^2$ as a function of pump power $P_0$ scaled in units of threshold power corresponding to Fig.~\ref{nonlinearities}. (d) Resonance of the enhancement factor as a function of delay $\tau$ corresponding to Fig.~\ref{decay-build-up} (e) Decrease of the enhancement factor as a function of maximum condensate blueshift $\text{max}[U(t)]$ over the pulse duration.}
    \label{fig.theory}
\end{figure}

The dynamics of a single excitation pulse are presented in Fig.~\ref{fig.theory}(a) for $P_0 = 1.5 P_\text{th}$. Notably, as the condensate forms the active reservoir $n_R$ depletes quickly which has some implications on the enhancement. Figures~\ref{fig.theory}(b) and~\ref{fig.theory}(c) show the time-integrated condensate density and enhancement as a function of pump power $P_0/P_\text{th}$, showing good agreement with the experimental results of Fig.~\ref{nonlinearities}. The non-smooth behavior in the data comes from the finite sample size of random noise realizations. 

Figure~\ref{fig.theory}(d) shows the resonant behavior of the enhancement as a function of delay, and how it diminishes at higher pump powers, in reasonable agreement with experiment. However, opposite to our measurements, the simulation depicts a slow build-up and fast decay of the enhancement. This behavior can be understood from the simple form of the stimulation term $R n_R$ and its corresponding gain-clamping term $R |\psi|^2$ in Eqs.~\eqref{eq:1} and~\eqref{eq:2}. The clamping term causes the reservoir to aggressively deplete and drop faster than the condensate [see Fig.~\ref{fig.theory}(a)]. This leads to the inverted shape of the enhancement factor in simulation seen in Fig.~\ref{fig.theory}(d) with respect to experiment. This behavior was found to be consistent over several investigated sets of parameters, which suggests that a more precise rate-equation model is needed to describe a slower depletion of the reservoir as the condensate builds-up. Notice also that the the simulated peak enhancement is located at $\tau \approx -6$\,ps, a shorter negative delay with respect to experiment, which can be understood from the fact that our pump term directly drives inactive excitons and skips charge-carrier generation, thus bypassing relaxation channels that would delay the build-up of the condensate and shift the enhancement peak further left. We leave this for future work on polariton cavities connected by feedback loops. 

Last, we explore the effects of the blueshift term $U(t)$ on the enhancement in Fig.~\ref{fig.theory}(e) by artificially increasing the value of $\hbar g_0$ linearly from $0 \to 50$ $\mu$eV $\mu$m$^{2}$. As expected, the enhancement drops as a function of the maximum blueshift over the pulse duration. The blueshift is found to mostly come from the reservoir populations $U \approx 2g_R(n_R + n_I)$. This decrease in enhancement is due to an increasing energy mismatch between the blueshifted feedback signal and the polariton state, i.e., the feedback signal has moved out of resonance. The result highlights the important role of polariton-exciton interactions in our system as a classical optical blockade.

\section{Conclusions}

We have performed an experimental and theoretical investigation on a nonresonantly pumped polariton cavity wherein the condensate emission is connected to a feedback loop. Our results demonstrate that by re-introducing polariton emission back into the pumped cavity through an external feedback arm has a significant effect on the condensate's emission characteristics, enhancing them by as much as $\approx 110\%$. This result has a promising impact on polariton-based strategies for unconventional~\cite{Kavokin_NatRevPhys2022} and neural-inspired computing~\cite{Opala_OptMatExp2023} that are dependent on the vivid input-output response of polariton condensates based on their strong optical nonlinearities. A classical rate-equation model describing a polariton condensate population responsible for the coherent light emission, coupled to populations of excitonic reservoirs, explains nearly all features of the study.

We believe that this method can be exploited in optical networks of condensate that are pumped with multiple localized spots focused on the cavity plane~\cite{Topfer_Optica2021, liang2025mirrormediatedlongrangecouplingrobust}. Different response can be triggered in each spatially separate condensate by feeding back into it weighted signals from other condensates (vector-matrix multiplication) from a previous pulse. Such a ``feedforward'' optical procedure can be implemented using reconfigurable holograms to help scale up polariton neurons~\cite{mirek2021neuromorphic, tyszka2023leaky} into functional networks.

\section{Acknowledgments}
M.F. acknowledges the project No. 2023/49/N/ST3/03595 funded by the Polish National Science Centre. H.S. acknowledges the project No. 2022/45/P/ST3/00467 co-funded by the Polish National Science Centre and the European Union Framework Programme for Research and Innovation Horizon 2020 under the Marie Skłodowska-Curie grant agreement No. 945339. A. O. acknowledges the National Science Center, Poland, project No. 2024/52/C/ST3/00324. B.P. acknowledges the project No. 2020/37/B/ST3/01657 funded by the Polish National Science Centre. This work was financed by the European Union EIC Pathfinder Challenges project ``Quantum Optical Networks based on Exciton-polaritons'' (Q-ONE, Id: 101115575). We are thankful to Michał Matuszewski for the supportive discussions.

\section{Appendix A: Experimental setup}

	\begin{figure}
		\centering
		\includegraphics[width=\linewidth]{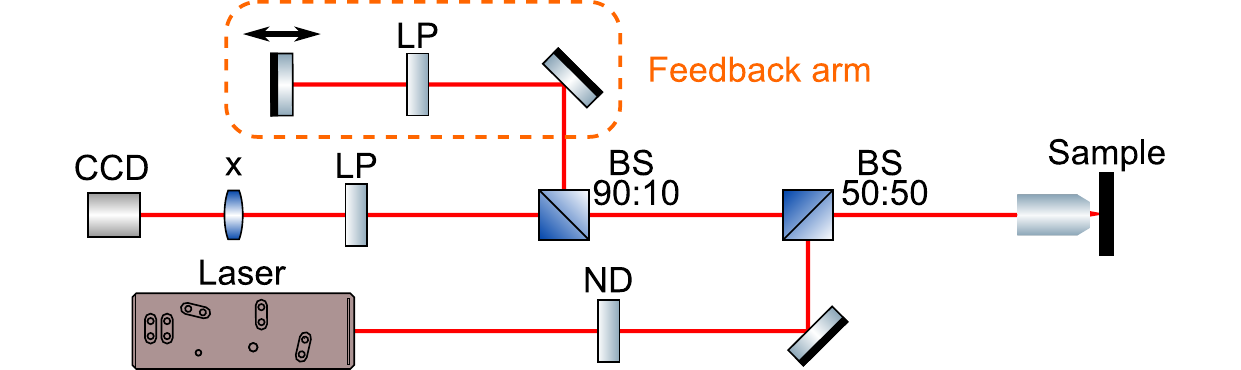}
		\caption{Experimental setup with additional feedback path, used during measurements.}
		\label{im:setup}
	\end{figure}

	\begin{figure*}
		\centering
		\includegraphics[width=\linewidth]{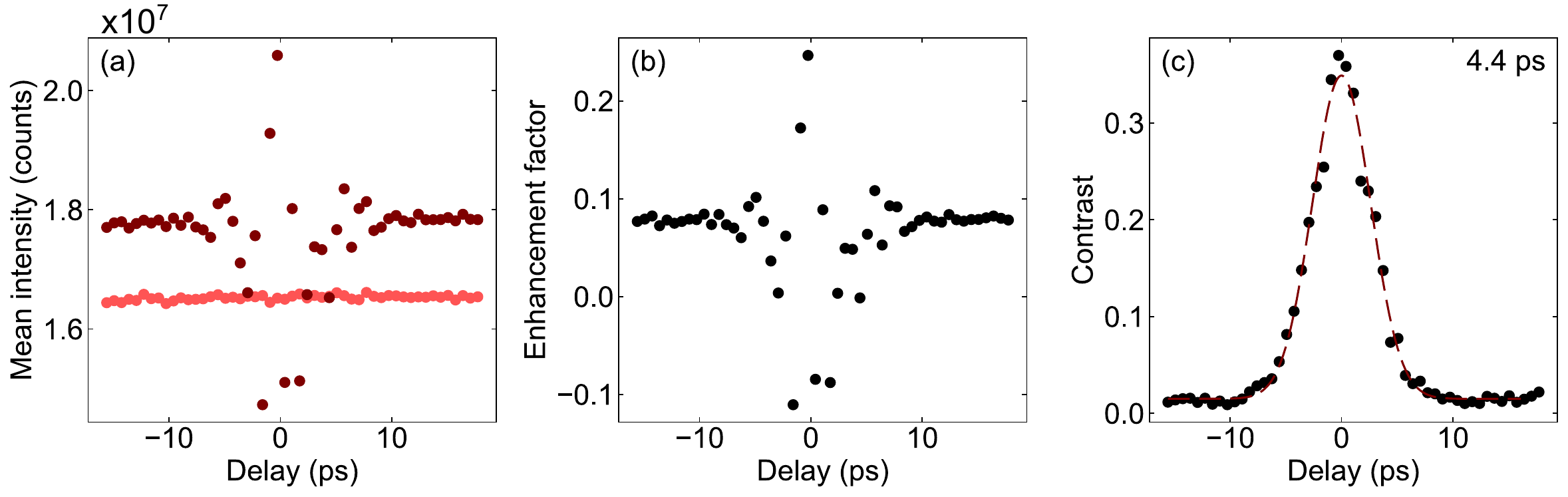}
		\caption{Panel (a) shows the signal intensity for different delays between the pulses. The bright red dots show the stability of the laser during the experiment, dark red dots correspond to the laser signal obtained with feedback arm. Each point corresponds to the average intensity collected from 100 frames with a locked feedback loop. Changes in signal intensity were greatest around zero delay. (b) Enhancement factor calculated for each delay based on average intensities. (c) Contrast calculated for each delay based on 100 maps measured with and without the feedback arm. The red dashed curve illustrates the Gaussian fit for a 4.4\,ps FWHM.}
		\label{im:laser-characterization}
	\end{figure*}
    
The scheme of the experimental setup is presented in Fig.~\ref{im:setup}. The sample was kept in the cryostat at liquid helium temperature and excited nonresonantly with a pulsed 4\,ps laser using an objective with a numerical aperture of 0.55. The laser energy was tuned to the first high-energetic Bragg minimum. The FWHM of the focused laser spot was of 1.7\,$\mu$m. The emission from the sample was collected using the same objective and then divided by a 90:10 (R:T) beam splitter~(BS) into two parts: feedback arm and the detection path. On the detection path, real space imaging was performed on a CCD camera. To cut off the nonresonant excitation laser and detect only the emission, long pass filter~(LP) was used. The feedback arm was used to take a portion of the cavity emission to be fed back into the cavity in order to provide an additional resonant enhancement of the condensation process in a subsequent nonresonant laser pulse. This is a type of a pump-probe setup which is similar to the polariton optical parametric oscillator~\cite{Savvidis_PRL2000, Stevenson_PRL2000}. The mirror on a translation stage in the feedback arm allowed us to variably control the time-delay between the signal coming from the feedback arm and the laser pulse exciting the sample. The distance between the mirror and the sample (around 3.94\,m) was set to match the repetition rate of the laser~(76\,MHz). To avoid exciting the sample with the laser beam reflected from the microcavity, the feedback arm also contained a long pass filter.

\section{Appendix B: Characterization of the laser}

The laser measurements allowing to determine zero time delay and the feedback arm split fraction are presented in Figure~\ref{im:laser-characterization}. The pulsed laser with an energy of 1.32\,pJ was reflected from the sample. The delay between pulses was varied and the signal observed on the CCD camera was compared both with and without the feedback arm. The experiment for each delay was repeated 100 times with an acquisition time of 100\,ms. In Fig.~\ref{im:laser-characterization}(a), the dark red color indicates how the laser intensity changed at different delays between pulses. The bright red points show the stability of the laser during the experiment. Each point corresponds to the average intensity collected from 100 frames with the feedback loop locked. These results show that the laser intensity was stable during the experiment, and the signal in the camera remained more or less the same. The signal observed for the laser with a feedback loop was qualitatively different. For delays greater than 10\,ps, a stable signal with higher intensity was observed than without the feedback loop. For shorter delays, interference between the pulses was observed, leading to significant changes in intensity over time. As a result, the mean intensity value fluctuated around \(1.8 \times 10^7\) counts. The changes were most significant around zero delay. Fig.~\ref{im:laser-characterization}(b) shows the quantitative contribution of the feedback loop to the total intensity observed on the CCD camera. The enhancement factor is below 0.1 for long delays. For delays close to zero, the feedback signal reflects the pattern observed in the overall intensity, which is the result of laser interference. Fig.~\ref{im:laser-characterization}(c) illustrates contrast determined from laser interference. Maximum contrast was observed at zero delay. A Gaussian function was fitted to the data and a FWHM of 4.4\,ps was obtained. The FWHM agrees with the laser pulse width, which is approximately 4\,ps.

\bibliography{bib}

\end{document}